\documentclass{PoS}

\title{Small--scale anisotropies of cosmic rays from relative diffusion }

\ShortTitle{Small--scale anisotropies of cosmic rays from relative diffusion }

\author{\speaker{Philipp Mertsch}\\
        Kavli Institute for Particle Astrophysics \& Cosmology, Menlo Park, CA 94025, USA\\
        E-mail: \email{pmertsch@stanford.edu}}

\author{Markus Ahlers\\
        WIPAC \& Department of Physics, University of Wisconsin--Madison, Madison, WI 53706, USA\\
        E-mail: \email{markus.ahlers@icecube.wisc.edu}}

\abstract{The arrival directions of multi--TeV cosmic rays show significant anisotropies at small angular scales. It has been argued that this small scale structure is reflecting the local, turbulent magnetic field in the presence of a global dipole anisotropy in cosmic rays as determined by diffusion. This effect is analogous to weak gravitational lensing of temperature fluctuations of the cosmic microwave background. We show that the non--trivial power spectrum in this setup can be related to the properties of relative diffusion and we study the convergence of the angular power spectrum to a steady--state as a function of backtracking time. We also determine the steady--state solution in an analytical approach based on a modified BGK ansatz. A rigorous mathematical treatment of the generation of small scale anisotropies will help in unraveling the structure of the local magnetic field through cosmic ray anisotropies.}

\FullConference{The 34th International Cosmic Ray Conference \\
		 30 July- 6 August, 2015\\
		 The Hague, The Netherlands}

\usepackage{amsmath}
\usepackage{aas_macros}

\begin{document}

%------------------------------------------------------------------------------------------------------------------------------------------------------------
\section{Introduction}

The arrival directions of cosmic rays (CRs) are highly isotropic, to about one part in $10^4$ at TeV--PeV energies. Given the discrete nature of their sources, this requires an efficient mechanism for isotropising their directions. Resonant interactions of CRs with turbulent magnetic fields, that lead to pitch--angle scattering and induce spatial diffusion, can provide this randomisation.

In a diffusive approximation to CR transport, the (microscopic) power spectrum of the turbulent magnetic field is related to the (macroscopic) transport parameters, like the pitch--angle or diffusion coefficients. The phase space density, averaged over a statistical ensemble of turbulent magnetic fields, is expanded into an isotropic part $f_0$ and anisotropic corrections $f_i(\mu)$~\cite{1966ApJ...146..480J}. (Here, $\mu$ is the pitch angle cosine.) Anticipating that the anisotropic corrections are much smaller than the isotropic part, the dipole anisotropy, defined by the first moment in pitch angle cosine, can be related to the spatial gradient of the isotropic part, $n = 4 \pi f_0$. The dipole term in the angular power spectrum of the relative intensity $\delta I\equiv(4\pi f-n)/n$ then relates to the gradient and the diffusion tensor ${\bf K}$ as,
\begin{equation}\label{eq:standardC1}
\frac{1}{4\pi}{C_1} = \left|\frac{{\bf K}\nabla n}{n}\right|^2\,.
\end{equation}
Another process contributing is the Compton--Getting effect~\cite{1935PhRv...47..817C}. Adopting a certain distribution of sources and a (semi--phenomenological), scalar diffusion tensor, it has been shown that the predicted dipole anisotropy is up to two orders of magnitude larger than the observed one~\cite{2005JPhG...31R..95H}. It has been suggested that intermittency effects in the source distribution~\cite{2006AdSpR..37.1909P,Blasi:2011fm,Pohl:2012xs} or (spatially dependent) anisotropic diffusion~\cite{Evoli:2012ha,2014ApJ...785..129K} can mitigate this discrepancy. An attractive solution is the combination of anisotropic diffusion with intermittency effects in the turbulent magnetic fields~\cite{Mertsch:2014cua}.

\begin{figure}[!bht]\centering
\includegraphics[width=0.5\linewidth]{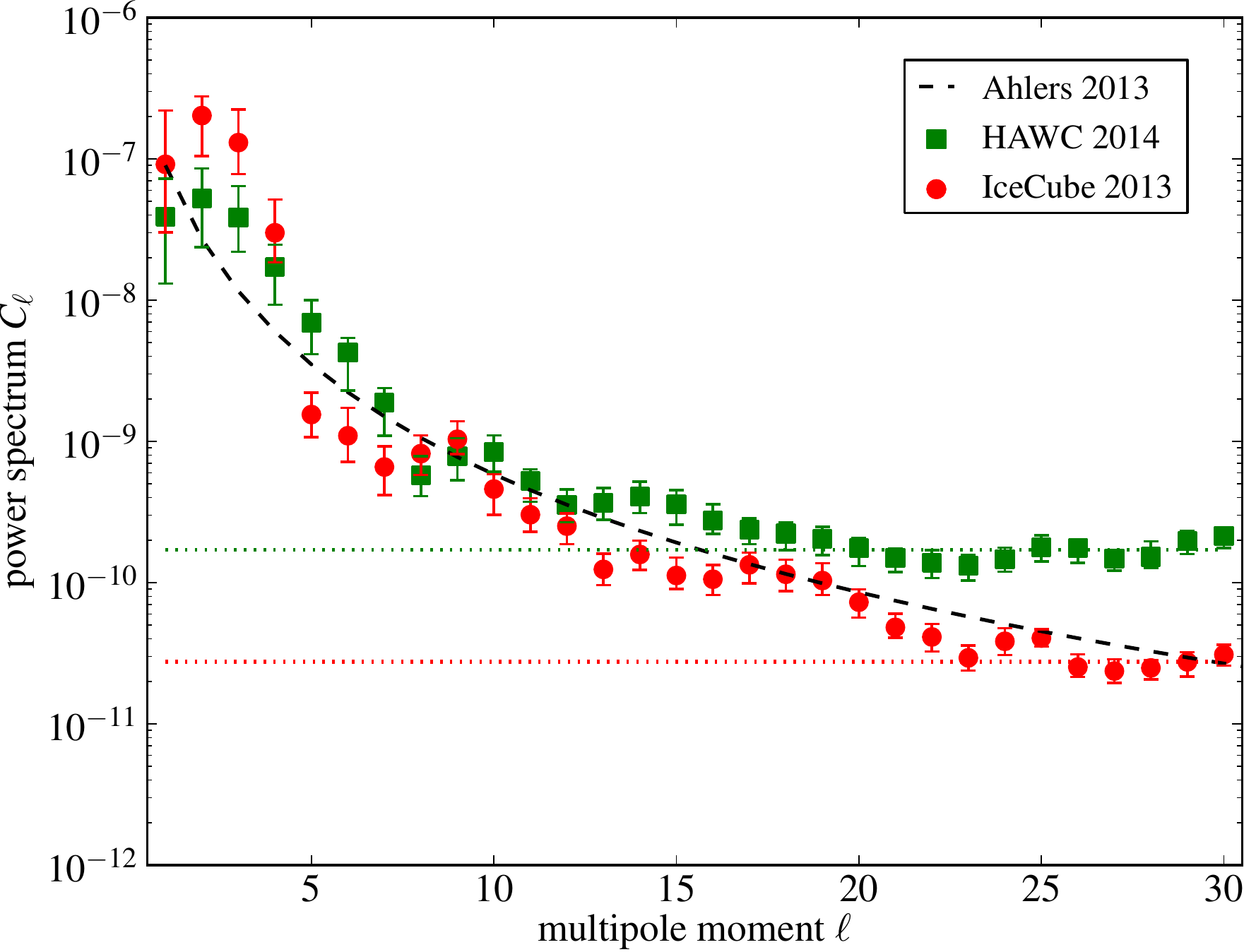}
\caption[]{Power spectrum of the CR arrival directions observed by IceCube~\cite{Aartsen:2013lla} and HAWC~\cite{Abeysekara:2014sna}. The horizontal dotted lines indicate the level of statistical noise at the level of $\mathcal{N} \simeq {f_{\rm sky}4\pi}/{N_{\rm CR}}$, where $f_{\rm sky}$ is the effective fraction of the sky that contains $N_{\rm CR}$ collected events. For IceCube we can estimate $f_{\rm sky}\simeq 0.3$ and $\mathcal{N}_{\rm IC} \simeq 2.5\times10^{-11}$ and for HAWC with $f_{\rm sky}\simeq 0.7$ we have $\mathcal{N}_{\rm HAWC} \simeq 1.8\times10^{-10}$. The dashed line shows the result of Ref.~\cite{Ahlers:2013ima}.}\label{fig1}
\end{figure}

Even more surprising, structure in the distribution of arrival directions has been detected at much smaller angular scales, first by MILAGRO~\cite{Abdo:2008kr}. IceCube~\cite{Aartsen:2013lla} and HAWC~\cite{Abeysekara:2014sna} have recently presented measurements of the angular power spectrum of the arrival directions, finding significant power down to angular scales of $\sim 5^\circ$ and $\sim 10^\circ$, respectively, see Fig.~\ref{fig1}. The appearance of structure on scales smaller than the dipole can also be understood in terms of intermittency in the turbulent magnetic fields~\cite{Giacinti:2011mz,Ahlers:2013ima}: Unlike the ensemble--averaged phase space density, the actual phase space density exhibits power on all scales. The particular realisation of the local turbulent magnetic field is thus imprinted on the arrival directions of cosmic rays. In Fig.~\ref{fig1}, we show the prediction of the ensemble--averaged angular power spectrum from a fully analytical computation of this effect~\cite{Ahlers:2013ima}.

In the following, we will support this idea with a combination of analytical and numerical arguments. Specifically, in Sec.~\ref{sec:relative} we will clarify the relation between the small--scale anisotropies and \emph{relative} diffusion of CRs. By numerically back--tracking CRs through turbulent magnetic fields, we will show in Sec.~\ref{sec:simulation} how the angular power spectrum develops in time and that convergence to a steady--state is reached for long enough backtracking times. In Sec.~\ref{sec:BGK-like}, we will describe the temporal evolution of the angular power spectrum, starting from Liouville's theorem and adopting a BGK--like~\cite{Bhatnagar:1954zz} ansatz. We will summarise and conclude in Sec.~\ref{sec:summary_conclusion}.

%------------------------------------------------------------------------------------------------------------------------------------------------------------
\section{Relative diffusion}
\label{sec:relative}

In the following, we study the small, stochastic deviations of the local phase space density $f$ from its ensemble average $\langle f \rangle$, $\delta f \equiv f - \langle f \rangle$. Given a quasi--stationary phase space density at time $t=-T$, $4\pi \langle f\rangle \simeq n -3\hat{\bf p}{\bf K}\nabla n$, the phase space density $f$ at time $t=0$ follows from Liouville's theorem,
\begin{equation}\label{eq:backtrack}
4\pi f \simeq 4\pi\delta{f}(-T) + n + ({\bf r}(-T)-3\hat{\bf p}(-T){\bf K})\nabla n\,,
\end{equation}
where ${\bf r}(-T)$ and ${\bf p}(-T)$ are the position and momentum of a CR particle at time $t=-T$. We can then compute the angular power spectrum at time $t=0$ which is defined as
\begin{equation}\label{eq:Celldef}
C_\ell = \frac{1}{4\pi}\int {\rm d}\hat{\bf p}_1 \int {\rm d}\hat{\bf p}_2 \, P_\ell(\hat{\bf p}_1  \hat{\bf p}_2) f_1f_2\,,
\end{equation}
where $\hat{\bf p}_{i}$ denote unit vectors and $P_\ell$ the usual Legendre polynomials. Here and in the following we use the abbreviation $f_i = f(t,{\bf r}_i,{\bf p}_i)$. We note that due to pitch--angle scattering, the term ${\bf r}(-T)\nabla n$ in eq.~\ref{eq:backtrack} will dominate over $-3\hat{\bf p}(-T){\bf K}\nabla n$ at late times. Furthermore, we assume that in the ensemble average and again at late times, fluctuations are uncorrelated, $\langle \delta f_1(-T) \delta f_2(-T) \rangle = 0$. We thus find for the ensemble--averaged angular power spectrum of the relative intensity $\delta I\equiv(4\pi f-n)/n$,
\begin{equation}\label{eq:Cell}
\frac{1}{4\pi}{\langle C_\ell\rangle}\simeq \int \frac{{\rm d}\hat{\bf p}_1}{4\pi} \int \frac{{\rm d}\hat{\bf p}_2}{4\pi}P_\ell(\hat{\bf p}_1  \hat{\bf p}_2) \lim_{T\to\infty}\langle {r}_{1i}(-T){r}_{2j}(-T)\rangle \frac{\partial_i n\partial_j n}{n^2} \, .
\end{equation}

This expression can be related to the symmetric part of the relative diffusion coefficient
\begin{equation}\label{eq:Krel}
\widetilde{K}^{\rm s}_{ij} =\int\frac{{\rm d}\hat{\bf p}_1}{4\pi}\int\frac{{\rm d}\hat{\bf p}_2}{4\pi} \lim_{T\to\infty} \frac{1}{4T}\big\langle \big\{r_{1i}(-T) - r_{2i}(-T) \big\} \big\{r_{1j}(-T) - r_{2j}(-T) \big\}\big\rangle\,,
\end{equation}
by noting that the sum over all multipoles $\ell \geq 0$ (i.e.\ the variance of the flux) reads
\begin{equation}\label{eq:id1}
\frac{1}{4\pi}\sum_{\ell\geq0}(2\ell+1){\langle C_\ell\rangle}(T) \simeq 2T{K}^{\rm s}_{ij}\frac{\partial_i n\partial_j n}{n^2}\,,
\end{equation}
where the symmetric part of the diffusion tensor ${K}^{\rm s}_{ij}$ is defined through $\langle{r}_{i}(-T){r}_{j}(-T)\rangle \to 2T{K}^{\rm s}_{ij}$ in the limit of large backtracking times $T$. The monopole on the other hand satisfies,
\begin{equation}\label{eq:C0}
\frac{1}{4\pi}{\langle C_0\rangle}(T) \simeq 2T\left({K}^{\rm s}_{ij}-\widetilde{K}^{\rm s}_{ij}\right)\frac{\partial_i n\partial_j n}{n^2}\,,
\end{equation}
such that the power in higher multipoles is
\begin{equation}\label{eq:id2}
\frac{1}{4\pi}\sum_{\ell\geq1}(2\ell+1){\langle C_\ell\rangle}(T) \simeq 2T \widetilde{K}^{\rm s}_{ij}\frac{\partial_i n\partial_j n}{n^2}\,.
\end{equation}

%------------------------------------------------------------------------------------------------------------------------------------------------------------
\section{Simulation}
\label{sec:simulation}

We now turn to a numerical simulation that backtracks CRs through a turbulent magnetic field. On top of a regular field ${\bf B}_0$, we have set up the turbulent field as the sum of $N$ harmonic modes $\delta{\bf B}({\bf x}) = \sum_{n=1}^N \delta {\bf B}_n \cos({\bf k}_n{\bf x}+\beta_n)$ with $N$ random phases $\beta_n$ and wave vectors ${\bf k}_n$, following Ref.~\cite{Giacalone:1999}. We choose the amplitudes to satisfy $\delta {\bf B}_n\perp{\bf k}_n$ and $|\delta {\bf B}_n|^2 \propto  k_n^3/(1+(k_nL_c)^\gamma)$ with coherence scale $L_c$, giving a Kolmogorow--type phenomenology for $\gamma = 11/3$. The level of turbulence is parametrised by $\sigma$ with $\sum |\delta {\bf B}_n|^2 = 2\sigma^2{\bf B}_0^2$. We sample the CR directions at time $t=0$, $\hat{\bf p}_i(0)$, isotropically on a \texttt{HEALPix}~\cite{Gorski:2004by} grid with resolution parameter $n_{\rm side}=32$, for a total of $N_{\rm pix}=12288$ trajectories per random field configuration, and we evaluate 120 such configurations. For the simulations, we choose $\sigma^2=1$, $\lambda_{\rm min}=0.01 \, L_c$ and $\lambda_{\rm max}=100 \, L_c$ with $k=2\pi/\lambda$. Whereas in the Galaxy with $L_c \sim 30 \, \text{pc}$ and $B_0 \sim 3 \, \mu\text{G}$ the gyroradius of $10 \, \text{TeV}$ CRs is much smaller than the coherence length, $r_L/L_c \sim 10^{-5}$, we here fix $r_L = 0.1 \, L_c$ to reach the (temporally) asymptotic regime within reasonable computing times.

\begin{figure}[!bht]\centering
\includegraphics[width=0.55\linewidth]{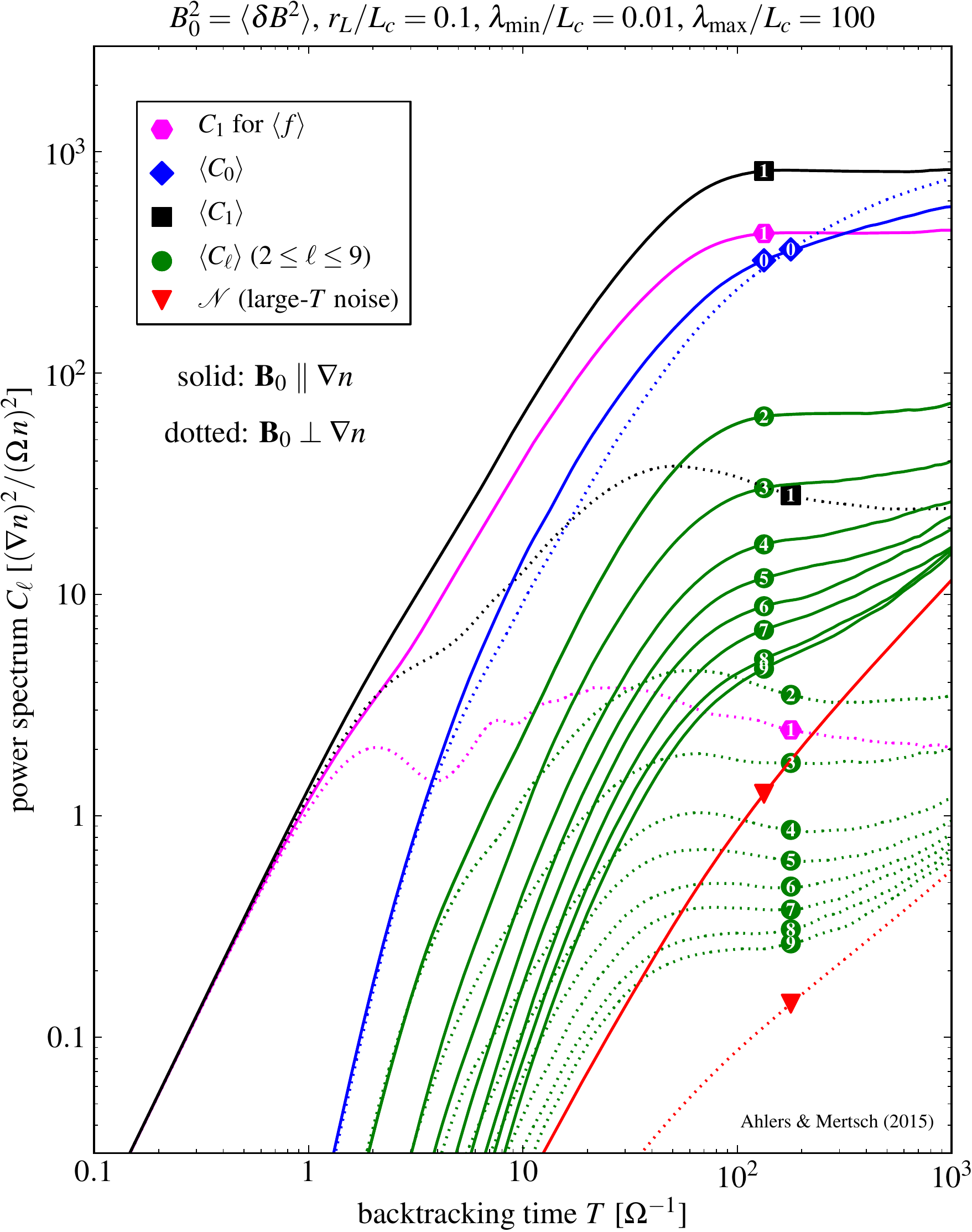}
\caption[]{The evolution of the ensemble-averaged power spectrum (\ref{eq:Cell}) for a CR gradient parallel (solid lines) and perpendicular (dotted lines) to the regular magnetic field and the 3D turbulence model discussed in the main text. We show results in terms of the dipole $\langle C_1\rangle$ (black), monopole $\langle C_0\rangle$ (blue) and medium-$\ell$ multipoles (green). We also show the asymptotic noise level (\ref{eq:noise}) (red) and the dipole prediction (\ref{eq:standardC1}) of standard diffusion (magenta) evaluated by the replacement $\langle{\rm r}_1{\rm r}_1\rangle \to \langle{\rm r}_1\rangle\langle{\rm r}_2\rangle$ in Eq.~(\ref{eq:Cell}).}\label{fig2}
\end{figure}

In Fig.~\ref{fig2}, we show the average angular power spectrum $C_\ell$ of the relative intensity $\delta I$ as a function of backtracking time $T$, obtained by multipole expansion with the \texttt{HEALPix} utilities. We show the results for a CR gradient parallel and perpendicular to the regular magnetic field ${\bf B}_0$ by the solid and dotted lines, respectively. Black lines denote the ensemble--averaged dipole, magenta lines show the standard dipole of the ensemble--averaged phase space density. As expected, the latter is smaller than the former. For $2\leq\ell\leq9$, the multipoles are shown by the green lines. As we are dealing with the angular power spectrum of the relative intensity, there is a small residual monopole which is shown by the blue lines.

At late times ($t \Omega \gtrsim 100$ for our parameters, $\Omega$ being the gyrofrequency), the angular power spectrum is exhibiting an asymptotic behaviour. This is reflecting the fact that the arrival directions measured at any one position are only sensitive to the particular realisation of the \emph{local} turbulent magnetic field. Memory of the anisotropies at even earlier times is lost. Note that the higher multipoles are more strongly effected by shot noise which is due to the finite number of trajectories and shown by the red lines in Fig.~\ref{fig2}. Again for large backtracking times, the shot noise can be estimated as
\begin{equation}\label{eq:noise}
\mathcal{N}
 \simeq \frac{4\pi}{N_{\rm pix}}2TK^{\rm s}_{ij}\frac{\partial_in\partial_jn}{n^2}\,.
\end{equation}

\begin{figure*}[!tbh]\centering
\includegraphics[width=0.48\linewidth]{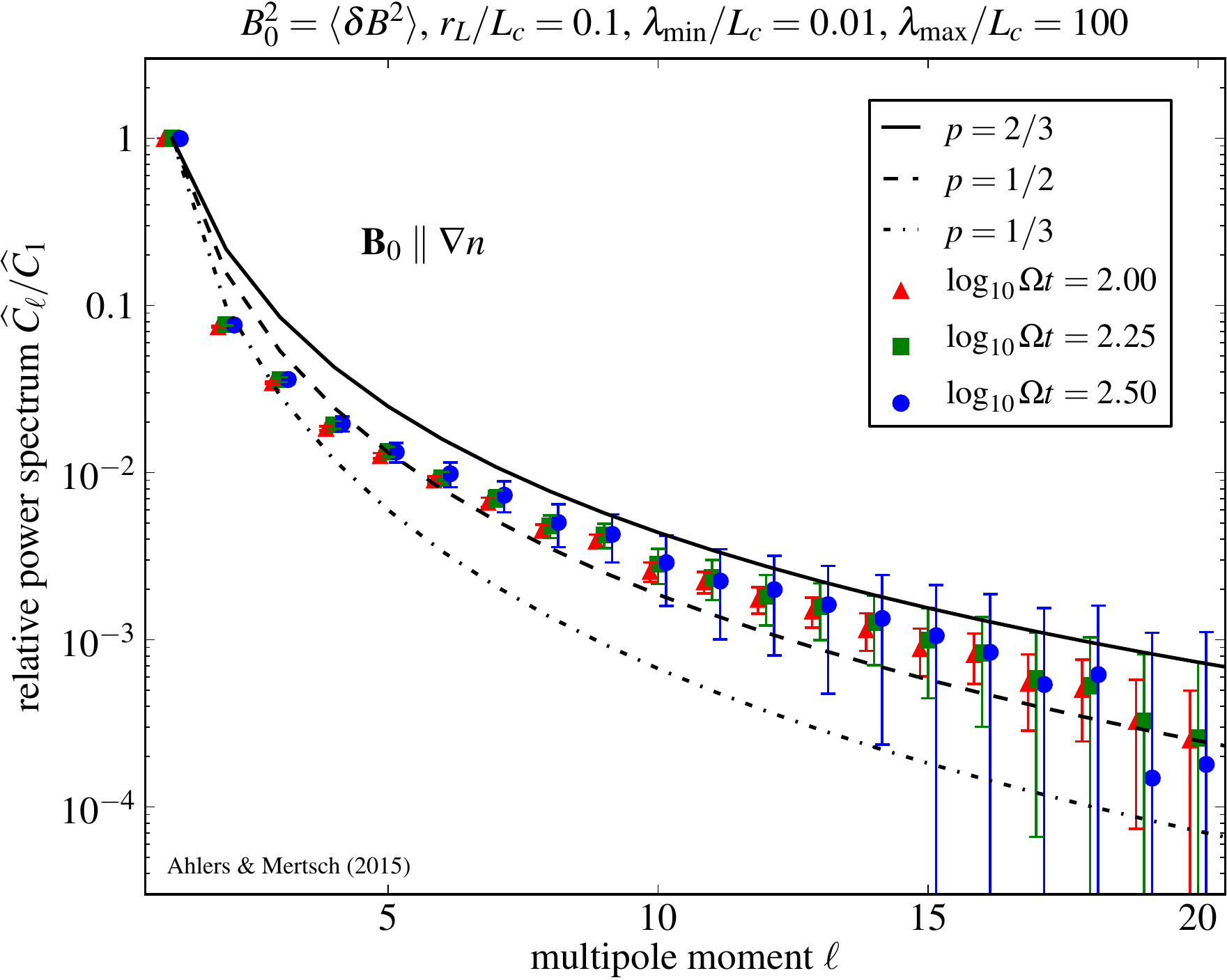}\hfill
\includegraphics[width=0.48\linewidth]{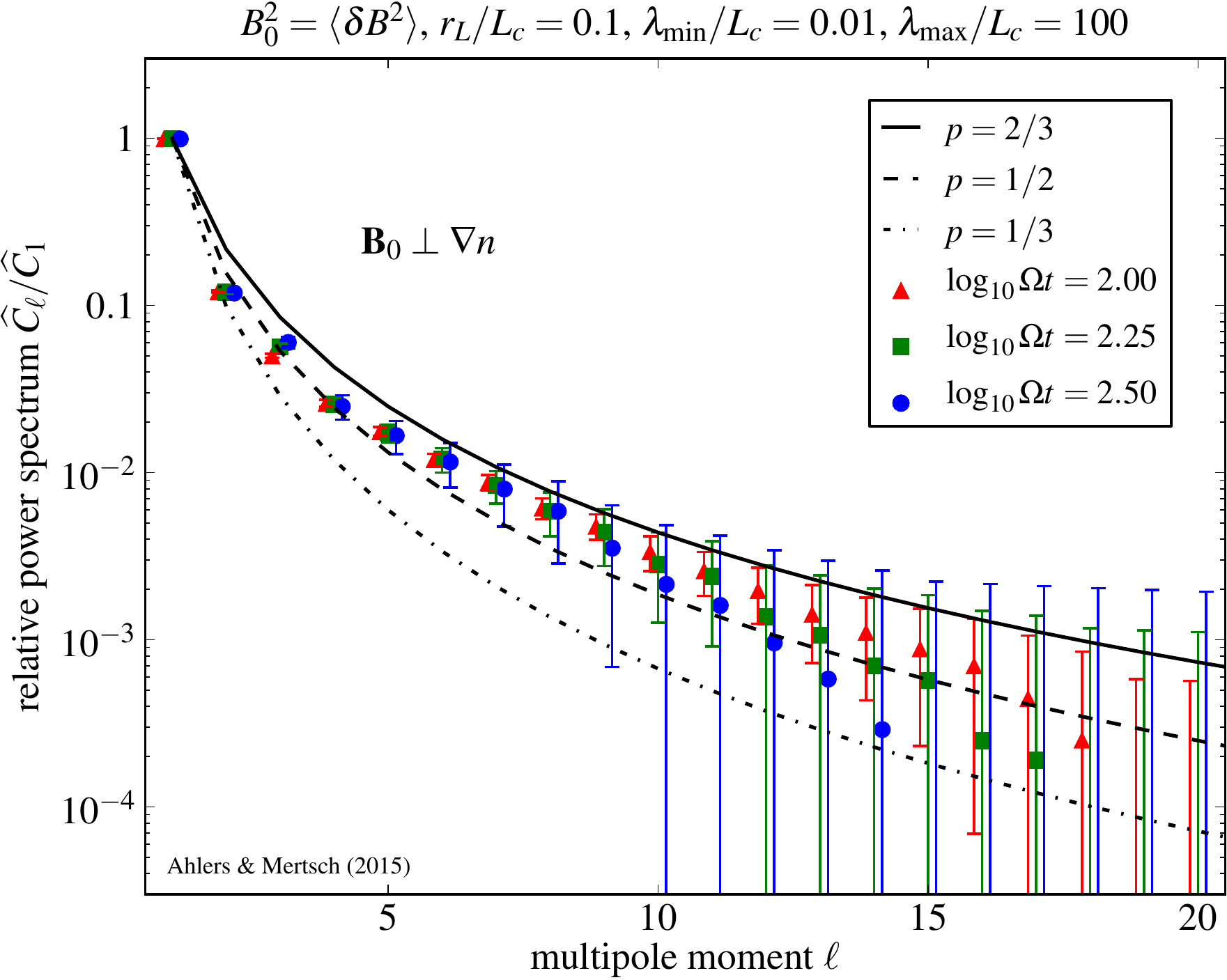}
\caption[]{The power spectrum estimator $\widehat{C}_\ell = \langle C_\ell\rangle -\mathcal{N}$ (normalized to $\widehat{C}_1$) for parallel (left plot) and perpendicular (right plot) CR gradient for the data shown in Fig.~\ref{fig2} at three different backtracking times. We also show the noise variance level of the data. The three lines correspond to the prediction of a relative scattering term $\nu_r(x) \propto (1-x)^p$ in Eq.~(\ref{eq:finalCl}).}\label{fig3}
\end{figure*}

In Fig.~\ref{fig3} we show the best estimator for the true power spectrum $\widehat{C}_\ell = \langle C_\ell\rangle-\mathcal{N}$ and its noise variance for three different times and for parallel and perpendicular CR gradients, respectively. For the limited number of trajectories considered here, it might be difficult to distinguish between noise variance and cosmic variance which is why we here show the former only. Note the concave shape of the angular power spectrum which is also seen in the experimental data (see Fig.~\ref{fig1}) and the convergence to the asymptotic form for late times.

%------------------------------------------------------------------------------------------------------------------------------------------------------------
\section{BGK--like ansatz}
\label{sec:BGK-like}

In the following, we derive the asymptotic angular power spectrum in a generalisation of the BGK~\cite{Bhatnagar:1954zz} ansatz used in diffusion theory. We start from the Boltzmann equation for the phase space density $f$, again split into an average and a fluctuating part, $f = \langle f\rangle +\delta{f}$. After ensemble averaging, the Boltzmann equation reads~\cite{JonesApJ1990}
\begin{equation}\label{eq:genBoltzmann}
\partial_t\langle f\rangle  + \hat{\bf p} \nabla_{\bf r}\langle f\rangle  -i\boldsymbol{\Omega}{\bf L}\langle f\rangle  = \left\langle i\boldsymbol{\omega}{\bf L}\delta{f}\right\rangle\,,
\end{equation}
where we have introduced the angular momentum operator ${\bf L} = -i{\bf p}\times\nabla_{{\bf p}}$ as well as the rotation vectors ${\boldsymbol \Omega} = e{\bf B}_0/p_0$ and ${\boldsymbol \omega} = e\delta{\bf B}/p_0$. In eq.~(\ref{eq:genBoltzmann}), the effect of the turbulent magnetic fields is contained in the r.h.s. collision term, $\left\langle i\boldsymbol{\omega}{\bf L}\delta{f}\right\rangle$. In standard diffusion theory, this term is replaced with a friction term following {\it Bhatnagar, Gross \& Krook}~\cite{Bhatnagar:1954zz} (BGK), that drives the ensemble--averaged phase space distribution to its isotropic mean $n$ with an effective relaxation rate $\nu$, i.e.
\begin{equation}\label{eq:BGKansatz}
\left\langle i\boldsymbol{\omega}{\bf L}\delta{f}\right\rangle \simeq -\nu\left(\langle f\rangle-\frac{n}{4\pi} \right) \, .
\end{equation}
In order to generalise the BGK ansatz, we note the relation to Brownian motion on the sphere~\cite{Yosida1949} where the diffusion equation reads $\partial_t f = -(\nu/2){\bf L}^2f$. Therefore, instead of eq.~\ref{eq:BGKansatz}, we write
\begin{equation}\label{eq:generalisedBGKansatz}
\left\langle i\boldsymbol{\omega}{\bf L}\delta{f}\right\rangle \simeq -\frac{\nu}{2} {\bf L}^2 \langle f \rangle \, .
\end{equation}

For the asymptotic angular power spectrum, we solve for the steady--state solution of
\begin{equation}\label{eq:CLevolution}
\partial_t\langle f_1f_2\rangle= \langle f_1\left(-\hat{\bf p}_2\nabla_{\bf r}+i\boldsymbol{\omega}_2{\bf L}+i\boldsymbol{\Omega}{\bf L}\right)f_2\rangle  + (1\leftrightarrow 2)\,.
\end{equation}
and in the spirit of eq.~\ref{eq:generalisedBGKansatz} we therefore make the ansatz
\begin{equation}\label{eq:BGK2}
\langle (i\boldsymbol{\omega}_1{\bf L}_1+i\boldsymbol{\omega}_2{\bf L}_2)f_1 f_2\rangle \simeq -\left[\nu_{\rm r}(x)\frac{{\bf L}_1^2+{\bf L}_2^2}{2} +\nu_{\rm c}(x){\bf J}^2\right]\langle f_1f_2\rangle\,,
\end{equation}
with $x=\hat{\bf p}_1  \hat{\bf p}_2$ and ${\bf J}={\bf L}_1+{\bf L}_2$. 
This term will lead to mixing and damping of multipoles. Here, we have defined the relative and correlated scattering rates, $\nu_{\rm r}$ and $\nu_{\rm c}$, that depend on the relative distance of the trajectories at early times, or equivalently on the opening angle between trajectories. Note that Eq.~(\ref{eq:BGK2}) is necessarily symmetric under interchange of ${\bf p}_1 \leftrightarrow {\bf p}_2$ as well as ${\bf p}_i\leftrightarrow -{\bf p}_i$ and is the most general linear approximation of all possible scalar combination of ${\bf L}_1$ and ${\bf L}_2$. In Eq.~(\ref{eq:CLevolution}), we also replace the gradient term in via $\langle f_1\hat{\bf p}_2\nabla f_2\rangle \simeq-3/(4\pi)^2\hat{\bf p}_1\nabla{n}\hat{\bf p}_2{\bf K}\nabla{n}$, assuming that the spatial dependence of $\delta f_i$ is small compared to $\langle f_i\rangle$ for $f_i = \langle f_i\rangle +\delta{f}_i$. 

One can show (see~\cite{Ahlers:2015dwa} for details), that the asymptotic angular power spectrum satisfies
\begin{equation}
Q_1\delta_{\ell1} = \sum_k\langle C_k\rangle k(k+1)\frac{2k+1}{2}\int {\rm d}x \, \nu_{\rm r}(x)P_\ell(x)P_k(x)\,,
\end{equation}
where the term $Q_1= {K}_{ij}{\partial_i n\partial_j n}/(6\pi)$ is sourcing the dipole. This can be inverted to give
\begin{equation}\label{eq:finalCl}
\langle C_\ell\rangle = \frac{3}{2}\frac{Q_1}{\ell(\ell+1)}\int\limits_{-1}^1{\rm d}x\frac{x\,P_\ell(x)}{\nu_{\rm r}(x)}\,.
\end{equation}

We model the unknown dependence of the relative scattering rate on the opening angle $x$ as $\nu_{\rm r}(x) \propto (1-x)^p$, with $0<p<1$. This is motivated by the fact that particles along the same trajectory never decorrelate and therefore $\nu_{\rm r}(1)=0$ whereas for $x<1$, $\nu_{\rm r}(x)$ must be positive but finite, and possibly monotonous. In Fig.~\ref{fig3}, we show the angular power spectrum for $p=1/3$, $1/2$ and $2/3$. The value $p=1/2$ in particular seems to be reproducing the simulated data rather well.

%------------------------------------------------------------------------------------------------------------------------------------------------------------
\section{Summary and conclusion}
\label{sec:summary_conclusion}

In summary, we have shown how the power in multipole $C_\ell$ with $\ell \geq 1$ relates to \emph{relative} diffusion of CRs. Furthermore, we have investigated the angular power spectrum of the relative intensity of CRs as a function of backtracking time, showing that at late times, the angular power spectrum converges to a steady state. This reflects the fact that the observed anisotropy reflects the \emph{local} realisation of the turbulent magnetic field and is independent of the anisotropy at much earlier times. Finally, we have also shown that the observed concave form of the angular power spectrum can be reproduced by adopting a BGK--like ansatz for the generalised collision term in the Boltzmann equation. Observations of the small--scale anisotropy therefore encode information about the \emph{actual} representation of the turbulent magnetic field in our Galactic neighbourhood. Such information could be harvested if the inverse problem of inferring the magnetic field from the distribution of arrival directions were tractable.

\bibliography{bibliography}

\end{document}